\documentclass[12pt]{article}
\usepackage[utf8]{inputenc}
\usepackage{lineno}
\usepackage{graphicx}
\usepackage{times}
\usepackage{geometry}
\usepackage[utf8]{inputenc}
\usepackage{enumitem,amssymb}
\usepackage{ragged2e}
\usepackage{graphicx}
\usepackage{sidecap}
\usepackage[super,sort&compress]{natbib}

\newlist{thematic}{itemize}{8}
\setlist[thematic]{label=$\square$}
\usepackage{pifont}
\newcommand{\cmark}{\ding{51}}%
\newcommand{\done}{\rlap{$\square$}{\raisebox{2pt}{\large\hspace{1pt}\cmark}}%
\hspace{-2.5pt}}

\usepackage[superscript,biblabel]{cite}
\usepackage{float}

\begin{document}

\raggedright
\huge
Astro2020 APC White Paper \linebreak
The Dark Energy Spectroscopic Instrument (DESI)
\linebreak
\normalsize

\noindent \textbf{Thematic Areas:} \hspace*{60pt} 
$\square$ Planetary Systems \hspace*{10pt} 
$\square$ Star and Planet Formation \hspace*{20pt}\linebreak
$\square$ Formation and Evolution of Compact Objects \hspace*{31pt} 
\done Cosmology and Fundamental Physics \linebreak
\done Stars and Stellar Evolution \hspace*{1pt} 
\done Resolved Stellar Populations and their Environments \hspace*{40pt} \linebreak
\done Galaxy Evolution   \hspace*{45pt} 
$\square$ Multi-Messenger Astronomy and Astrophysics \hspace*{65pt} \linebreak
  
\vspace{-0.1in}
\textbf{Principal Authors:}
\linebreak
Michael E. Levi (Lawrence Berkeley National Laboratory) \\ \&  
Lori E. Allen (National Optical Astronomy Observatory) 
\linebreak 
\textbf{Email:} melevi@lbl.gov, lallen@noao.edu 


\textbf{Co-authors:}
Anand Raichoor (EPFL, Switzerland), Charles Baltay (Yale University), Segev BenZvi (University of Rochester), Florian Beutler (University of Portsmouth, UK), Adam Bolton (NOAO), Francisco J. Castander (IEEC, Spain), Chia-Hsun Chuang (KIPAC), Andrew Cooper (National Tsing Hua University, Taiwan), Jean-Gabriel Cuby (Aix-Marseille University, France), Arjun Dey (NOAO), Daniel Eisenstein (Harvard University), Xiaohui Fan (University of Arizona), Brenna Flaugher (FNAL), Carlos Frenk (Durham University, UK), Alma X. Gonz\'alez-Morales (Universidad de Guanajuato, M\'exico), Or Graur (CfA), Julien Guy (LBNL), Salman Habib (ANL), Klaus Honscheid (Ohio State University), Stephanie Juneau (NOAO), Jean-Paul Kneib (EPFL, Switzerland), Ofer Lahav (UCL, UK), Dustin Lang (Perimter Institute, Canada), Alexie Leauthaud (UC Santa Cruz), Betta Lusso (Durham University, UK), Axel de la Macorra (UNAM, Mexico), Marc Manera (IFAE, Spain), Paul Martini (Ohio State University), Shude Mao (Tsinghua University, China), Jeffrey A. Newman (University of Pittsburgh), Nathalie Palanque-Delabrouille (CEA, France), Will J. Percival (University of Waterloo, Canada), Carlos Allende Prieto (IAC, Spain), Constance M. Rockosi (UC Santa Cruz), Vanina Ruhlmann-Kleider (CEA, France), David Schlegel (LBNL), Hee-Jong Seo (Ohio University), Yong-Seon Song (KASI, South Korea), Greg Tarl\'e (University of Michigan), Risa Wechsler (Stanford University), David Weinberg (Ohio State University), (Christophe Y\`eche (CEA, France), Ying Zu (Shanghai Jiao Tong University, China)\\


\smallskip
\noindent\textbf{Abstract:}
We present the status of the Dark Energy Spectroscopic
Instrument (DESI) and its plans and opportunities for the coming
decade.  DESI construction and its initial five years of operations are an approved experiment of the U.S.\ Department of Energy and is summarized here as context for the Astro2020 panel. Beyond 2025, DESI will require new funding to continue operations.  We expect that DESI will remain one of the world's best facilities for wide-field spectroscopy throughout the decade. More about the DESI instrument and survey can be found at 
https://www.desi.lbl.gov.

\pagebreak



\newcommand\todo[1]{{\bf TODO: #1}}
\newcommand\num{{\bf NUM}}
\newcommand\inline[1]{\noindent{\bf #1: \hspace{0pt}}}



\section{An Overview of DESI: 2020-2025}

DESI is an ambitious multi-fiber optical spectrograph sited on the
Kitt Peak National Observatory Mayall 4m telescope, funded to conduct
a Stage IV spectroscopic dark energy experiment.  DESI featuers
5000 robotically positioned fibers in an 8 deg$^2$ focal plane,
feeding a bank of 10 triple-arm spectrographs that measure the full
bandpass from 360 nm to 980 nm at spectral resolution of 2000 in
the UV and over 4000 in the red and IR (Martini et al. 2018).  DESI is designed for efficient operations and
exceptionally high throughput, anticipated to peak at over 50\% 
from the top of the atmosphere to detected photons, not counting obscuration of the telescope or aperture loss from
the $1.5''$ diameter fibers.  More information is in Table \ref{tab:desi}.

\begin{figure}[b!]
\centering
\includegraphics[width=0.95\textwidth]{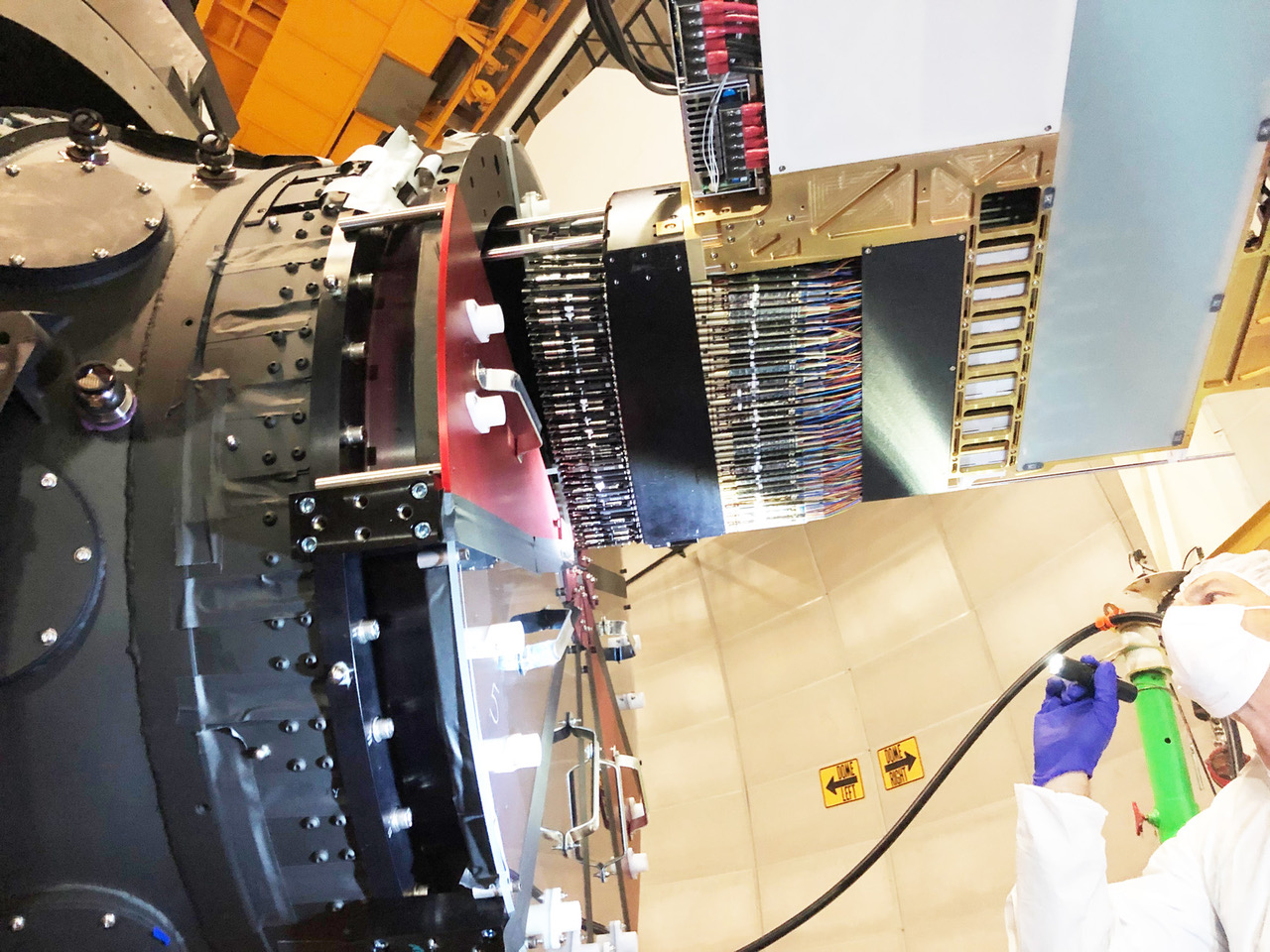}
\caption{\label{fig:petal}The first of 10 focal plate petals installed at the Mayall prime focus (June 26, 2019).}
\end{figure}

\begin{figure}[b!]
\centering
\includegraphics[width=0.95\textwidth]{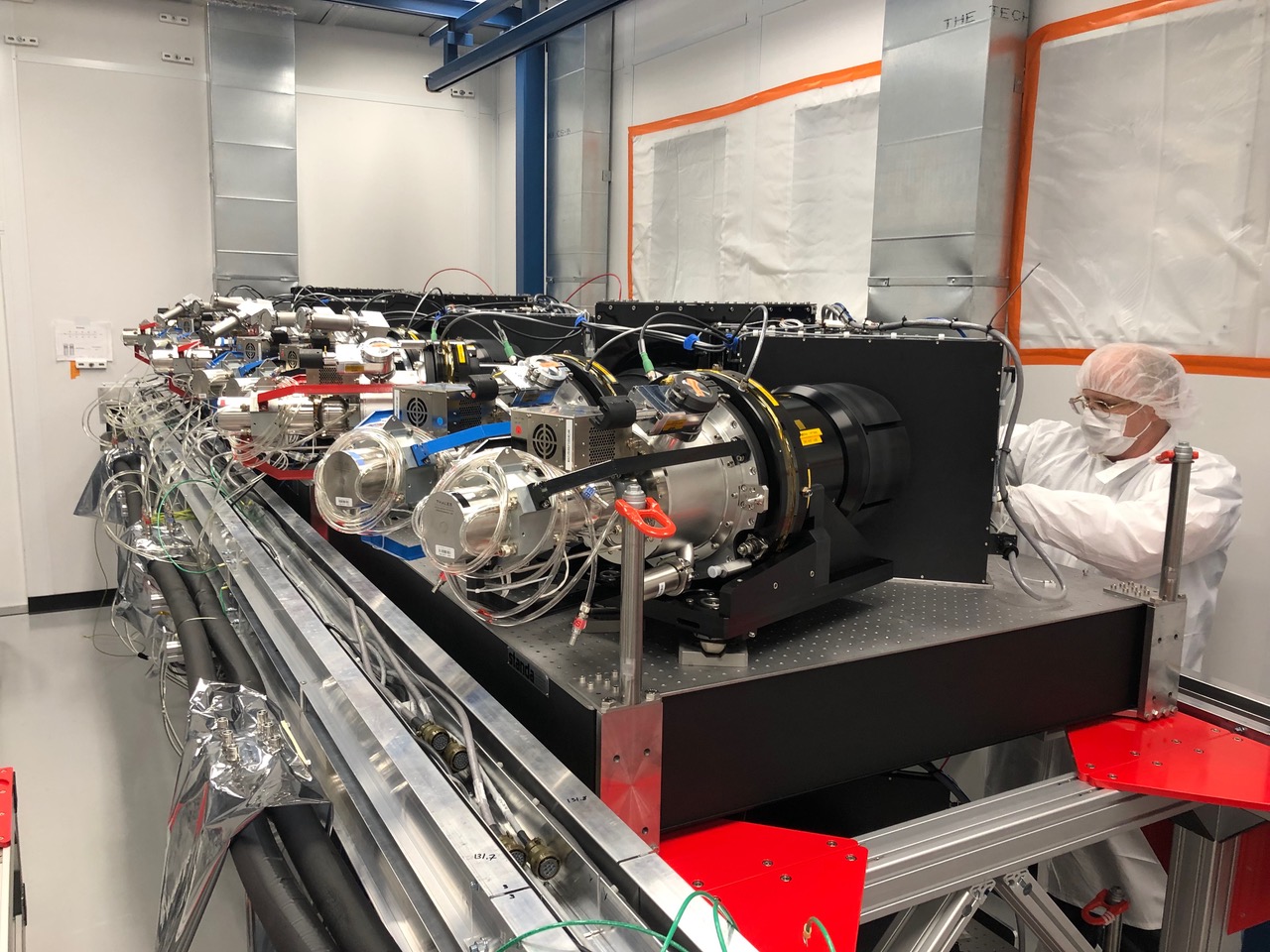}
\caption{\label{fig:spect}Six of the ten 3-armed DESI spectrographs, installed in their thermal enclosure.}
\end{figure}

\begin{figure}[b!]
\centering
\includegraphics[width=0.75\textwidth]{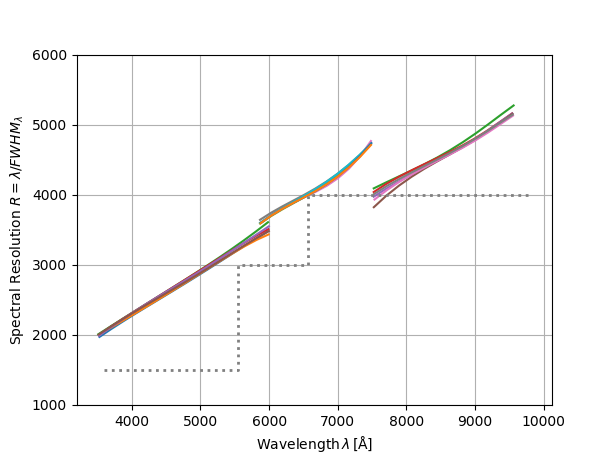}
\caption{\label{fig:res}Measured resolution of the six installed spectrographs.  Dotted lines are the system requirements.  The as-built results match exquisitely 
to the modeled performance.}
\end{figure}

As of this writing in July 2019, DESI construction is nearly complete
and the instrument is being installed at the Mayall telescope.  The
new prime-focus corrector was operated on sky in April/May 2019 and
confirmed to produce sharp images.  All ten petals of the robotic positioners and
all fibers have been constructed; these are being installed on the
telescope in July (Figure \ref{fig:petal}).
Six of the ten spectrographs
are installed and entering off-sky commissioning (Figure \ref{fig:spect} and \ref{fig:res}); the other four
should arrive in fall 2019.  We anticipate spectroscopic first-light in
October 2019, with commissioning running through January 2020.  The
collaboration will then operate a 4-month Survey Validation program
in spring 2020 and begin the 5-year survey in summer 2020. 

\begin{table}[p!]
\centering
\includegraphics[width=0.95\textwidth]{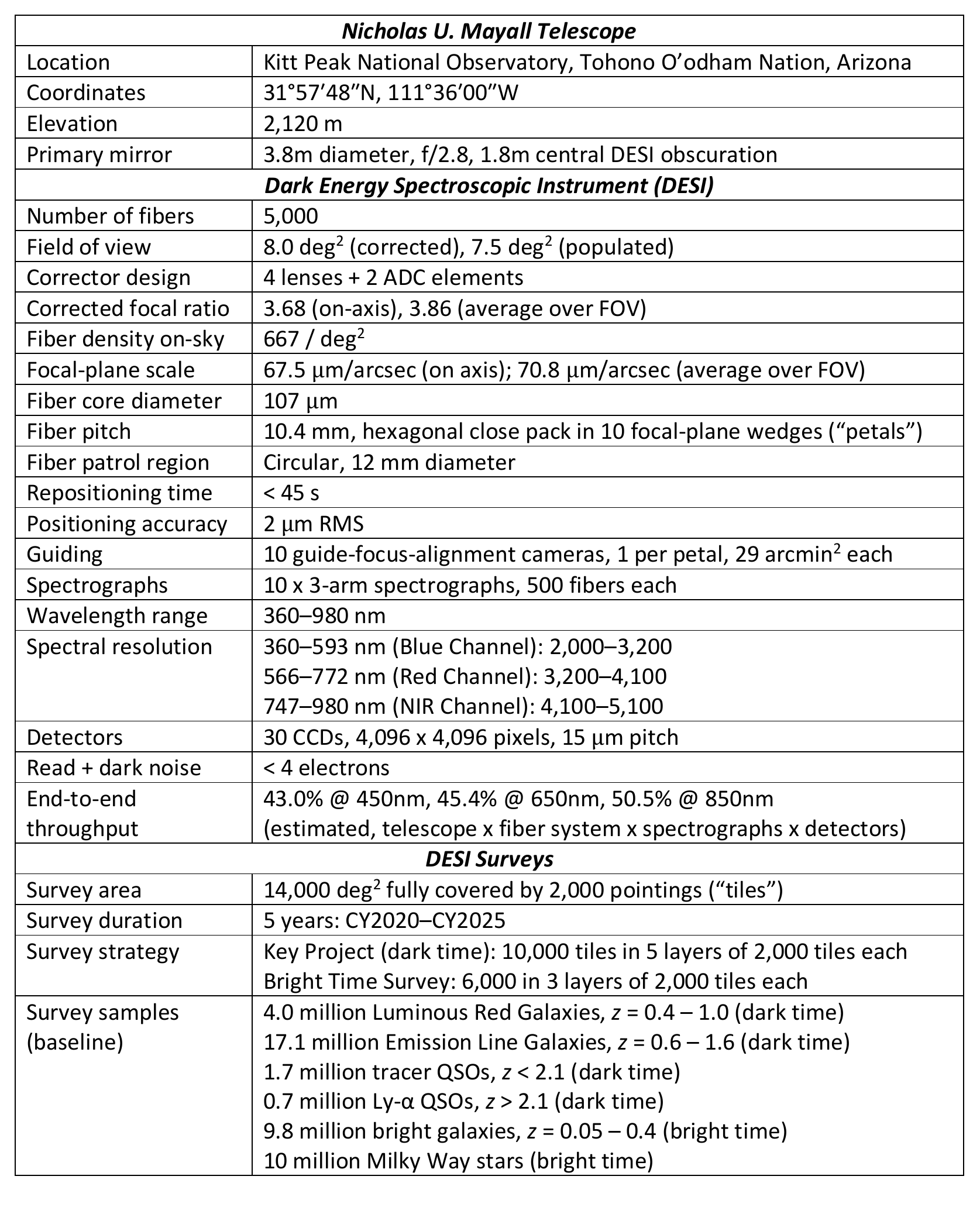}
\caption{\label{tab:desi}DESI at a glance.}
\end{table}

\inline{Key Science Goals}
The DESI Collaboration will use this facility to conduct a 5-year
survey of galaxies and quasars, covering 14,000 deg$^2$ and yielding
34 million redshifts.  The mission-need science of this survey is the study
of dark energy through the measurement of the cosmic distance scale
with the baryon acoustic peak method as a standard ruler and through the study of the
growth of structure with redshift-space distortions.  The survey
will further allow measurement of other cosmological quantities,
such as neutrino mass and primordial non-Gaussianity, as well as
studies of galaxies, quasars, and stars.

The DESI survey uses a sequence of target classes to map the large-scale
structure of the Universe from redshift 0 to 3.5 (Aghamousa et al. 2016).
In dark and grey time, DESI will utilize quasars, emission-line galaxies,
and luminous red galaxies.
Over 4M luminous red galaxy sample will cover $0.3<z<1$, including coverage
to $z\sim0.8$ at a density twice that of SDSS-III BOSS.  
The emission-line galaxy sample is the largest set, 18M, covering 
$0.6<z<1.6$ and providing the majority of the distance scale
precision.  2.4M quasars selected from their WISE excess will 
extend the map.  Importantly, these will yield Lyman $\alpha$ forest
measurements along 600K lines-of-sight, from which we will measure
the acoustic oscillations at $z>2$.

In bright time, DESI will conduct a flux-limited survey of 10M
galaxies to $r\approx19.5$, with a median redshift around 0.2.
This will allow dense sampling of a volume over 10 times that
of the SDSS MAIN and 2dF GRS surveys, which we expect will spur
development of cosmological probes of the non-linear regime of
structure formation.

In addition to extragalactic targets, DESI will observe many millions
of stars.  About 10M stars at $16<G<19$ will fill unused fibers in
the bright time program, and we will conduct a backup program of
brighter stars when observing conditions (clouds, moon, and/or
seeing) prevent useful data from being collected on extragalactic
targets.  Because DESI is a bench-mounted spectrograph with sub-degree
temperature stability, we anticipate velocity precision to $\sim1$~km/s. 

The DESI Collaboration plans for release of annual data sets,
including survey selection functions and mock catalogs suitable for
clustering analyses, following completion of its cosmology key
projects.  

\begin{figure}[b!]
\centering
\includegraphics[width=0.95\textwidth]{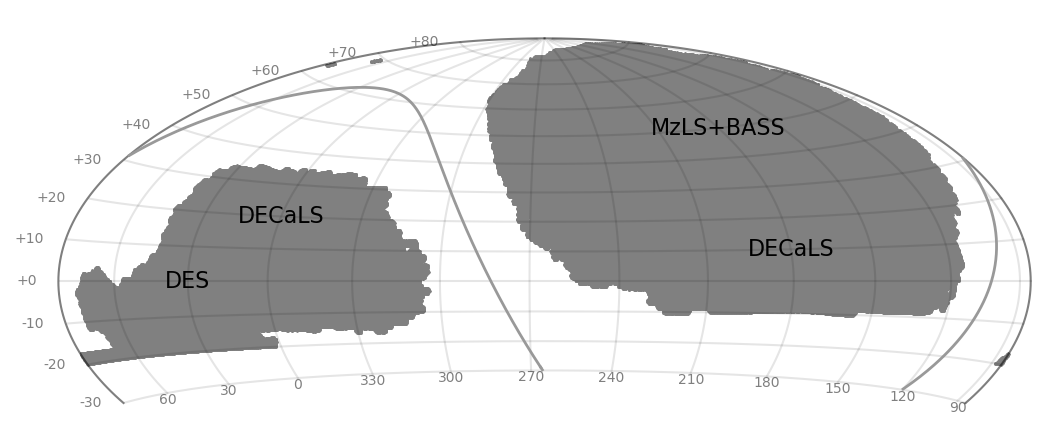}
\caption{\label{fig:foot}The planned DESI survey footprint is shown by the shaded area; it is built on several existing imaging surveys and extends as far south as $\delta=-20^\circ$ in the SGC and $-10^\circ$ in the NGC.}
\end{figure}

\inline{Target Selection}
In preparation of target selection for DESI, the Collaboration has
played a leading role in the execution of the Legacy Survey imaging
program, using nearly 1000 nights on the Blanco, Mayall, and Bok
telescopes to image 15,000 deg$^2$ to $g=24$, $r=23.4$, and $z=22.5$
depth, co-reduced with 5 years of WISE satellite imaging (Dey et al. 2019).  This is
the deepest coverage of the full high-latitude sky in the Northern
hemisphere (Figure \ref{fig:foot}).  The imaging data and catalogs have had 8 data releases,
available at {\tt http://legacysurvey.org}, the last of which reaches over 19,000 deg$^2$ by inclusion
of the 5-year Dark Energy Survey. Hence, DESI has already provided
an extensive data product for the general astronomical community.



As regards this first five-year survey with DESI, we stress the 
opportunity of this U.S.-led project to conduct cutting-edge 
dark energy research, both in its own right and in coordination
with optical, millimeter, and X-ray imaging data sets.  As 
well illustrated by the SDSS, the combination of spectroscopy and
imaging unlocks a wide range of applications.  


\inline{Organization}
DESI is being built by the DESI Collaboration with primary funding
from the U.S.\ Department of Energy, and additional funding from
the National Science Foundation, the Science and Technologies Facilities Council of the United Kingdom, the Gordon 
and Betty Moore Foundation, the Heising-Simons Foundation, the National Council of Science and Technology of Mexico, the French Alternative Energies and Atomic Energy Commission (CEA), and by the DESI 
Member Institutions.  The DESI Collaboration
currently has more than 600 total members from 79 institutions from 13 countries around the world.
Of those, $\sim$200 are senior members, and $\sim$400 are early career scientists. 

With DESI, the Mayall telescope is dedicated to this single instrument
configuration (unlike DECam on the Blanco, which could also mount
a secondary mirror).  The DOE Office of Science High-Energy Physics
division will be the primary funder of the DESI survey, including
the operation of the Mayall telescope through the 5-year survey.
The survey will utilize at least the darkest 21 nights per lunation,
plus engineering time, and potentially may use all of the telescope time.

\inline{Cost}
The DESI construction project has cost \$75M, with \$56M
from the DOE and the balance by other partners and institutional
buy-ins. Survey operations are budgeted at $\sim\!\$12$M/year, split approximately one-third site operations, and the rest supporting the instrument, the survey planning, and data processing and catalog creation. 


\clearpage
\section{Beyond 2025 with DESI}

Beyond the end of the first 5-year survey, DESI will remain a
state-of-the-art facility for wide-field surveys.  New commitments
for funding will be required.  But given the time scale to construct
a facility more powerful than DESI and with no such project yet
approved\footnote{We note that DESI (under the previous name BigBOSS) was identified 
in New Worlds New Horizons as an exemplar of the MidScale Innovation 
Program.  Despite timely agency support (CD-0 approved in 2012 and CD-2 in 2015), non-federal funding to conduct long-lead procurements, and no major programmatic or technical interruptions, we will be in operations in 2020.  We believe this is indicative of what projects of this complexity require, even in good outcomes!} 
we expect that the ground-based facility landscape in the
second half of the 2020's will look much like the first.  See Table \ref{tab:facilities} for a summary.  We note that not only is DESI at the forefront of this generation, without it, the U.S.\ community will not have a facility to compete with ESO and Subaru.
Further, we note that while Euclid and WFIRST will offer space-based 
platforms for slitless IR spectroscopy, optical spectroscopy remains 
a highly efficient way to get redshifts both at $z<1.5$ and $z>2$.


\begin{table}[b!]
    \centering
    \begin{tabular}{cccccc}\hline
        Name & Telescope & \# Fibers & FOV (deg$^2$) & Bandpass (nm) & Resolution \\
        \hline
        DESI & Mayall 4-m    & 5000 & 8 & 360--980  & mid \\
        PFS  & Subaru 8-m    & 2400 & 1.5 & 380--1260 & mid \\
        4MOST & VISTA 4-m    & 2436 & 5 & 370--950  & mid \& high \\
        WEAVE & WHT 4-m      &  960 & 3 & 370--960  & mid \& high \\
        SDSS-V & Sloan \& DuPont 2.5-m & 1000 & 7 & 360--1700 & mid (opt) \& high (IR) \\
        \hline
    \end{tabular}
    \caption{A brief comparison of multi-fiber facilities under construction.  Mid-resolution is typically a few thousand; high-resolution is typically around 20K, but for a more limited bandpass.  DESI will offer the highest multiplex and largest field of view of these next-generation facilities; only PFS has more instantaneous light-gathering power, but it is not a dedicated platform. Of the current generation of facilities, LAMOST is operating 4000 fibers in a 20 deg$^2$ field of view, but with performance limited to bright galaxies and stars.}
    \label{tab:facilities}
\end{table}

\inline{Key Science Goals}
We anticipate that a second phase of DESI will continue to offer
exciting survey opportunities.  Certainly we will not have exhausted
the supply of plausible targets on the sky.  Imaging surveys from
HSC, LSST, Euclid, SphereX, WFIRST, eROSITA, and others will yield improved
isolation of valuable targets over areas of thousands to tens of thousands of deg$^2$.  Spectroscopy can provide the key 
leverage to realize the science potential of these candidates, whether for redshifts or for more detailed
characterization.  DESI's combination of field of view, 
multiplex, throughput, and resolution makes it a great
complement to the coming generation of imaging surveys.

There are at least 5 fertile areas of potential targets for such a 
survey:

1) High-redshift emission-line targets: improved selection of $1<z<1.6$
emission-line galaxies from deeper imaging; selection of Ly$\alpha$
emission candidates from deep imaging in the blue; or follow-up of 
low-quality emission-line candidates from Euclid and SPHEREx.  Such a survey
would increase the sampling of the large volume available at 
higher redshift.

2) Increased depth and sampling in the Lyman-$\alpha$ forest, 
reobserving known targets and adding fainter candidates from 
deeper imaging.

3) A high-density galaxy survey at $z<1$, building on the DESI
bright galaxy survey.  These candidates are readily identified, but
a high-density sample with precise spectroscopic redshifts would allow identification of groups and
redshift-space distortions in the non-linear regime within the cosmic
acceleration epoch.

4) A high-multiplex survey of the Milky Way, with $O(100)$ million
stars, to yield radial velocities and stellar abundances to pair
with the exquisite Gaia astrometry.  The rapid reconfiguration time
of DESI ($<2$ minutes) makes short exposures an effective strategy.

5) Time-domain spectroscopy and transient host spectroscopy, building
on SDSS-V and time-domain imaging surveys such as ZTF, LSST, and TESS.

\inline{Technical Drivers}
While the DESI instrument could continue to be usefully operated in the same 
configuration as the pre-2025 phase, there may be opportunities for 
augmentations.  Notably, the spectrographs are modular and could be 
altered or replaced, subject to cost and space constraints, if the adopted science goals called for it.

\inline{Organization, Status, and Schedule}
A science collaboration for post-2025 operations has not yet been
formed, but we expect that many of the current participants would be
interested in continuing.  The Mayall telescope remains property
of the National Science Foundation, while the DESI equipment 
is DOE property.

We expect that planning for such surveys will pick up speed in 2021 with
the arrival of early DESI data, which will solidify the on-sky
performance and give a tactile sense of the target selections.

We note that there has been mention of the idea of moving DESI
south to the Blanco in 2025. On the plus side, such a move would 
increase sky overlap with LSST. On the down side, it is an expensive 
proposition: it has taken over 1.5 years to install DESI at the Mayall, and 
much of that work would need to re-occur.  We therefore expect that a
move would result in substantial downtime for both telescopes,
along with financial cost. Such a decision will require a detailed cost-benefit 
analysis. Given the large amount of near-equatorial
sky visible jointly from Arizona and Chile, we suspect that many
post-2025 survey options could be well performed without a move,
potentially in collaboration with an instrument of similar or even
lesser capability in the south.

\inline{Cost Estimate}
The budget for a second phase of DESI operations would depend on the
survey choices made as well as on the assessment of costs of ongoing
instrument support, presumably informed by experience in the coming 
years of operations.  However, the cost is likely $O(\$10-15)$M/year (inclusive), 
comparable to those of other mid-scale facilities that deliver highly
processed data products.  Hence, we expect ongoing operations to fall
in the Medium class of ground-based activities.

In conclusion, we expect that the Mayall telescope with DESI will
remain a world-class facility for high-multiplex optical
mid-resolution spectroscopy in the latter half of this decade, 
offering the U.S.\ the opportunity to continue its leadership
in spectroscopic wide-field surveys.


\bigskip
\textbf{References} \\
Dey, A., et al.,
''Overview of the DESI Legacy Imaging Surveys,''
ApJ, 157, 168 (2019)

Martini, P., et al., 
``Overview of the Dark Energy Spectroscopic Instrument,''
SPIE,  107021F, (2018).

DESI Collaboration, Aghamousa, A., et al., 
``The DESI Experiment Part I: Science,Targeting, and Survey Design,''
arXiv:1611.00036 (2016)

DESI Collaboration, Aghamousa, A., et al., 
``The DESI Experiment Part II: Instrument Design,''
arXiv:1611.00037 (2016)

\end{document}